\documentclass[%
 reprint,
 amsmath,amssymb,
 aps,
]{revtex4-2}

\usepackage{graphicx}
\usepackage{dcolumn}
\usepackage{bm}
\usepackage{hyperref}
\usepackage{float}
\begin{document}
\preprint{APS/123-QED}
\title{Machine Learning based Glitch Veto \\ for inspiral binary merger signals using \\Linear Chirp Transform}
\author{N.Arutkeerthi}
\email{n.arutkeerthi@ph.iitr.ac.in}
%\altaffiliation{Indian Institute of Technology,Roorkee}
\author{Xiyuan Li}
\email{xli522@uwo.ca}
%\altaffiliation{Western University,Ontario}
\author{Cindy Cui}
\email{ccui62@uwo.ca}
%\altaffiliation{Western University,Ontario}
\author{SR Valluri}
\email{valluri@uwo.ca}
%\altaffiliation{Western University,Ontario}
\affiliation{Indian Institute of Technology,Roorkee}
\affiliation{Western University,Ontario}
\date{\today}
\begin{abstract}
Transient non-Gaussian noise artifacts commonly known as glitches remain a major challenge in gravitational wave (GW) detection because they can mimic genuine compact binary coalescence signals and increase the false-alarm rate of detection pipelines. Accurate discrimination between astrophysical signals and instrumental glitches is therefore essential for improving the reliability of GW observations. In this work, we investigate the Linear Chirp Transform (LCT) as a feature extraction technique for glitch classification. Unlike the conventional Fourier transform, the LCT incorporates an additional chirp-rate parameter $\gamma$, enabling improved representation of signals with time-varying frequencies. Applying the LCT to GW strain time series produces three-dimensional chirp-volume spectrograms spanning time, frequency and chirp-rate dimensions, providing richer information than conventional time-frequency spectrograms. The dataset consists of confirmed compact binary coalescence events and glitch samples from the O1-O4 observing runs of the LIGO detectors at Hanford and Livingston. For classification, we employ a hybrid deep learning architecture combining convolutional neural networks (CNNs), gated recurrent units (GRUs) and an attention mechanism. The CNN layers extract local spectro-temporal features, the GRUs model correlations across chirp-rate slices and attention pooling highlights the most informative regions. The proposed framework achieves high classification performance on training and validation datasets, demonstrating that chirp-domain representations provide highly discriminative information for distinguishing merger signals from glitches. These results highlight the potential of combining chirp-based signal processing with deep learning to improve glitch mitigation in current and future GW observatories.
\end{abstract}
\maketitle
\section{\label{sec:level1} Introduction}

Gravitational waves have become a critical area of research for validating Einstein's theory of General Relativity, with Laser Interferometer Gravitational-Wave Observatory ( LIGO ) and VIRGO ( European Gravitational-Wave Observatory ) confirming the first direct detection in 2015 \cite{r4,r5}. With additions of other detectors like the Japanese gravitational wave KAGRA also known as Large-Scale Cryogenic Wave Telescope ; upcoming additions like LIGO India ; the advent of third-generation gravitational wave detectors like Einstein Telescope, Cosmic Explorer; upcoming Laser Interferometer Space Antenna ( LISA ) and other upcoming projects like Taiji, TianQin and Deci-Hertz Interferometer Gravitational Observatory ( DECIGO ) \cite{r35,r36,r37,r38,r39,r40,r41,r42} would expand our current knowledge on gravitational waves and their sources with more precise detections. These waves are ripples in the fabric of spacetime, produced when massive objects like binary black holes or neutron stars \cite{r32,r33} experience strong gravitational interactions and move rapidly. The most common sources of detectable gravitational waves are binary systems of black holes or neutron stars, which emit signals as they spiral toward each other and eventually merge- referred to as inspiral-merger signals \cite{r10}.

The detection of the gravitational wave merger signals are in general done by templated or non-templated searches \cite{r7,r8}.
Non-templated searches are based on the detection of a sudden chirp or increase in frequency similar to the simulated waveforms of binary merger signals. Templated search, as the name suggests, is searching via the correlation of a given signal to existing modelled templates. For templated search, the most widely used and accurate search is matched filtering  \cite{r9}. Here we have a catalog of templates of waveforms that are simulated by changing the mass, distance from the detector's line of sight, spin, etc, of the binary merger system but mainly the masses $( m_1,\; m_2 )$ in order to get an accurate template. A given signal is matched with this template by correlation, and if it matches beyond a threshold - it is considered a candidate event, after which it has to be checked to see if it is a genuine merger signal. This has been the most reliable technique so far for detecting an inspiral binary merger waveform from a given signal.

However, since the first detection researchers have encountered signals that mimic these merger events, known as glitches \cite{r6}. These detectors are very sensitive and detect signals and monitor changes in length down to \(10^{-19} \;  \text{m}/\sqrt{\text{Hz}}\) variations \cite{r11} which is comparable to dimensions of the size of small atoms hence our detectors have to be very precise and sensitive. The presence of glitches that are much higher in noise and such transient artifacts mimicking the merger signals reduce the sensitivity of detectors significantly. Glitches are caused by various factors - such as environmental disturbances, instrumental noise, or even cosmic rays, and can obscure or imitate genuine gravitational wave signals. This presents a significant challenge for data analysis, as glitches must be accurately identified and filtered out to avoid misinterpreting noise as real astrophysical events. We want to eradicate every possible situation of false alarms to have the most accurate signal detection possible. Therefore every candidate event must go through a vetoing technique, where the glitches are ruled out and classified away from actual binary merger signals. We are taking into account those glitches that do bypass the matched filtering stage due to their close resemblance to the merger waveforms.

The paper will be organized as follows: In Section 2, we will discuss briefly the existing glitch veto methods and their limitations. Continuing in Section 3, we will introduce the JCTFT technique briefly and explain why we choose chirp transform for our inspiral signals. In Section 4, we will discuss data preparation and processing, followed by Section 5 exploring Principal Component Analysis to check the separability without further advanced Deep Learning networks and the neural network architecture used for classification. Finally, we will discuss the Network model, results  and conclusions in Section 6, Section 7 and Section 8 respectively.

\section{\label{sec:level2}Existing Glitch-Veto Methods}
\subsection{Traditional Chi-Square Veto}
Initially, the Chi-square method \cite{r3,r59} was used to distinguish glitches from real signals by comparing the orthogonal projections of a signal in a parameter space. The Chi-square ( $\chi^2$ ) method is a statistical technique that measures how well an observed data set matches an expected model or distribution. In the context of glitch detection in gravitational wave signals, the Chi-square method involves calculating the discrepancies between the observed signal and a theoretical template across different segments of the signal.
For each segment, the signal is projected onto an orthogonal parameter space defined by the physical characteristics of the system, such as frequency, phase or amplitude. By comparing these projections to the expected values based on models of merger signals, the Chi-square statistic can quantify how closely the signal aligns with a real gravitational wave event. The sum of squared differences between the observed and expected projections across all segments is computed to yield the Chi-square value.
If the signal significantly deviates from the model in some regions, as is often the case with glitches - the Chi-square value increases flagging the signal as anomalous. Conversely, a low Chi-square value suggests the signal closely matches the expected pattern, making it more likely to be a genuine gravitational wave event. 
However, the challenge with this approach lies in the diversity of glitches - since there are around 22 distinct classes of glitches \cite{r45,r62,r64} %add citations- there are more than 23 classes!
with different characteristics, the parameter space for each glitch type needs to be carefully tailored, making the method computationally expensive.  Moreover, the Chi-square method is not completely infallible - glitches with certain characteristics that can still slip through the veto mechanism and lead to false positives in the detection process. A different 
parameter space is required for different glitch types. %insert citations for the same with statistical ideas
Despite these limitations, the core idea remains foundational in many veto techniques:  finding a parameter space where the behavior of glitches and merger signals diverges, allowing for their separation.

\subsection{Recent Glitch-veto techniques: Unphysical templates}

In response to the limitations of traditional glitch-classification techniques, researchers have explored alternative veto strategies based on unphysical waveform templates \cite{r2}. The central idea is to identify regions of the parameter space that are unlikely to correspond to physically realistic compact binary coalescence signals and use them to distinguish glitches from genuine gravitational-wave events. Such approaches attempt to construct parameter spaces in which glitches and merger signals occupy distinct regions, thereby improving their separability.

One successful implementation of this idea introduced chirp-time parameters, $\tau_0$ and $\tau_1$, derived from post-Newtonian waveform models \cite{r2}. These parameters are related to the frequency evolution of the signal and provide an alternative representation of the binary system that can be used to identify physically and unphysically motivated regions. Similar approaches based on chirp-mass parameterizations have also been investigated \cite{r12}. Template matching within these parameter spaces is commonly performed using optimization techniques such as Particle Swarm Optimization (PSO) \cite{r29}.

These methods have demonstrated excellent performance, achieving glitch-veto efficiencies approaching 99.9\% when combined with additional consistency tests involving signal-to-noise ratio and time-of-arrival information \cite{r2}. However, they remain dependent on waveform approximations and predefined parameter spaces, which may not fully capture the diversity of observed glitches or the complete range of astrophysical signals \cite{r28}. Furthermore, the effectiveness of these methods can be influenced by the choice of optimization strategy and detector noise characteristics.

In parallel, several machine-learning-based approaches have been developed for glitch classification, including Generative Adversarial Networks (GANs), similarity-learning methods and detector-characterization frameworks \cite{r55,r56,r57}. These methods learn discriminative features directly from the data and have shown promising performance across a wide range of glitch morphologies.

Our work is motivated by the broader objective of identifying representations in which merger signals and glitches exhibit clear separability. Rather than constructing a low-dimensional parameter space from waveform models, we investigate the Linear Chirp Transform (LCT) as a feature-extraction tool. The LCT generates a three-dimensional representation of the signal in chirp-rate, frequency and time, allowing us to analyze the evolution of signal structure across the chirp domain. By combining these chirp-volume spectrograms with deep learning techniques, we seek to develop a robust classification framework that can complement existing veto strategies for candidate gravitational-wave events.
\section{\label{sec:lev3}Joint Chirp Rate Time Frequency Transform ( JCTFT )}
In previous research \cite{r1}, the Joint Chirp Rate Time Frequency Transform ( JCTFT ) technique was introduced as a novel method for analyzing gravitational wave signals. The JCTFT extends the conventional Fourier transform by incorporating a linear chirp rate,  denoted by $\gamma$ which represents the rate of change of frequency over time ( with units of $T^{-2}$ ). Specifically, a linear chirp signal's frequency can be modeled as $f(t)=f_0 + \gamma t$, where the frequency changes linearly with time at the rate $\gamma$. This allows the JCTFT to capture the time-frequency behavior of signals that exhibit frequency evolution, which is typical for many astrophysical processes including inspiral-merger signals.

The importance of the chirp rate in gravitational wave signals, particularly those from inspiral - merger events,  cannot be overstated. Such signals evolve in frequency as compact objects spiral toward each other before merging, resulting in a rapidly increasing frequency near the merger. In essence, the JCTFT correlates the signal not only with frequency but also with the chirp rate by introducing a $\gamma t^2$ term in the phase of the transform. This addition enables the detection of signals with varying frequencies by accounting for the rate at which the frequency changes. While the frequency model of inspiral-merger signals is complex and typically nonlinear, a linear chirp serves as a first approximation for improving the response to these signals.

Thus, with the JCTFT we obtain a 3D spectrogram (time, frequency and chirp rate $\gamma$) instead of a traditional 2D time-frequency spectrogram. This added dimension provides richer information, potentially revealing subtle features in the signal that may be lost in a purely time-frequency analysis. Figure 1 illustrates the 3D spectrogram we get for a given signal. Mathematically, the JCTFT can be expressed as an extension of the standard Fourier transform:
\begin{figure*}[htbp]
    \centering
    \includegraphics[width=\textwidth]{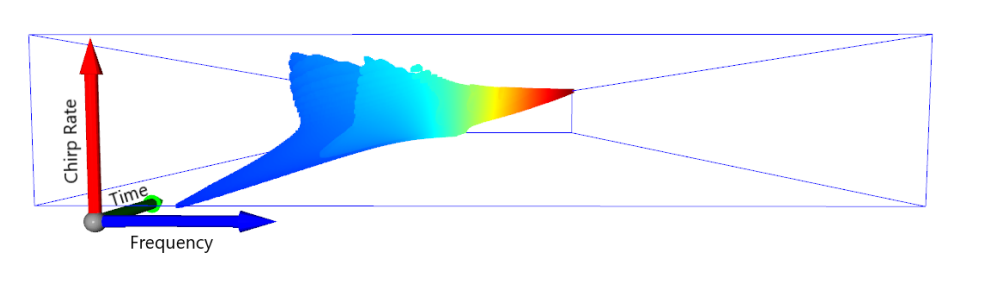}\hfill
    \caption{3D spectrogram output of a clean merger signal strain visualized. X. Li, M. Houde, and S. Valluri et., al. \cite{r1} \newline Notation: The Time axis is in the direction into the page}
\end{figure*}

\begin{equation} H(\Omega) = \int_{-\infty}^{\infty} h(t) e^{-i2\pi\Omega t} , dt \end{equation}

Equation (1) represents the traditional Fourier transform \cite{r30,r31}, where the signal $h(t)$ is correlated with a phase function of frequency $\Omega$. This gives insight into the power distribution in the frequency domain. However, this approach is limited when dealing with signals whose frequency evolves over time. To address this, the JCTFT introduces a chirp term:

\begin{equation} H(\Omega, \gamma) = \int_{-\infty}^{\infty} h(t) e^{-i2\pi(\Omega t + \gamma t^2)} , dt \end{equation}

Equation (2) represents the Linear Chirp Transform ( LCT ), a generalization of the Fourier transform that introduces the $\gamma t^2$ term in the phase. This additional term accounts for the linearly increasing frequency ( i.e., the chirp ) at the rate $\gamma$. When differentiating the phase, it becomes evident that the frequency of the signal increases linearly over time, making the JCTFT highly suitable for signals like gravitational waves, where the frequency evolution is a key feature.

In the process of applying the JCTFT to a signal $h(t)$, discretization \cite{r23} becomes crucial for practical analysis. The discretized JCTFT can be represented as a convolution between the Fourier transform of $h(t)$ and the Fourier transform of the window function that incorporates the chirp $\gamma t^2$ term. Specifically, for the product of two functions in the time domain, the convolution theorem allows us to take the convolution of the Fourier transforms of the individual functions shown in Equations 3 and 4. We choose a Gaussian window function for this convolution although other window functions can be selected for different purposes, such as to enhance time or frequency resolution.

{\small\begin{equation}
H(\gamma, \tau, \Omega) = \mathcal{F}\{f(t) g_c(t - \tau, \gamma, \Omega)\} = \mathcal{F}\{f(t)\} \ast \mathcal{F}\{g_c(t - \tau, \gamma, \Omega)\}
\end{equation}}

{\small\begin{equation}
= \mathcal{F}\{f(t)\} \ast \mathcal{F}\{g_c(t, \gamma, \Omega)\} e^{-i2\pi\Omega\tau}
\end{equation}}

In equation 5, $g_c$ represents the Gaussian window function, and $\tau$ denotes the time shift.\\
The window function $g_c$ is defined as follows:

\begin{equation} g_c(\gamma, t - \tau, \Omega) = \frac{\left|\Omega_0 + \mu\Omega\right|}{\sqrt{2\pi \left(1 + \frac{t - \tau}{|t - \tau|^2}\right)}} e^{-(t - \tau)^2 \left(\frac{(\Omega_0 + \mu\Omega)^2}{2} + i2\pi\gamma\right)} \end{equation}
This window function \cite{r20} is key to extracting meaningful frequency and chirp rate information from a given segment of the signal. A detailed analysis of different window functions and their impact on chirp transform performance can be found in \cite{r1}. In our work, we opt for the Gaussian window due to its well-known properties in time-frequency analysis, balancing the trade-off between time and frequency localization. Especially for signals with chirp, it smoothens out local regions of the signal, effectively preventing any spectral leakage. The window function adjusts well according to the frequency and chirp variations locally.

Here, $g_c$ incorporates the chirp term and normalization factors, ensuring that the window function is appropriately scaled. This function resembles sliding functions used in time-frequency analysis techniques like the S-Transform \cite{r21}. A significant advantage of the JCTFT lies in its ability to provide a more sensitive response to signals with chirping behavior compared to traditional Fourier-based techniques, which are less effective in capturing non-stationary features.

The Fourier transform of the window function $g_c$ has an exact solution, derived as follows:
{\small\begin{equation} G_c(\gamma, \Omega, \alpha) = \frac{|\Omega_0 + \mu\Omega|}{\sqrt{2\pi}} \left( e^{-\pi^2 \alpha^2 / z} \left( \frac{\sqrt{\pi}}{2 \sqrt{z}} \left( 1 - \operatorname{erf}\left( \frac{i \pi \alpha}{\sqrt{z}} \right) \right) \right) \right) \end{equation}}
In this expression, $G_c(\gamma, \Omega, \alpha)$ represents the Fourier transform of the window function, without any time shift, where $\alpha$ is the shift in frequency. The availability of an exact solution simplifies the process of computing the JCTFT, as the Fourier transform of the signal $h(t)$ can be convolved with this pre-computed solution, thereby enabling efficient computation. It is possible to have higher chirp terms ( say $t^3, t^4...$ and even fractional chirp term $t^{-7/6}$, etc. ) in such a way that it exactly matches the order of our inspiral merger signal's waveforms but they would not have an exact solution and give us a nice discretized version like the Linear Chirp transform. There is a huge possibility of inferences that could open up if we could derive the analytical solutions of the same.
But for now in \cite{r1}, it was demonstrated that gravitational wave signals from inspiral-merger events, which exhibit an increasing frequency as the objects approach each other, are well-suited for analysis using the JCTFT. This is because the positive chirp of these signals naturally aligns with the chirp dimension in the JCTFT, enhancing signal detection even in low signal-to-noise ratio ( SNR ) conditions. Empirical tests with SNRs as low as 5, up to 15, consistently showed that the JCTFT provided better performance in detecting these signals than the Q-Transform for signals with lower SNRs

\begin{figure}[htbp] \centering \fbox{\includegraphics[width=0.95\linewidth]{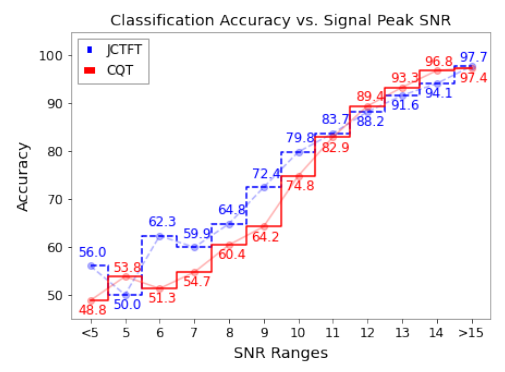}} \caption{Comparison of Detection accuracy by InceptionV3 Network on JCTFT v/s Q-Transform. X. Li, M. Houde, and S. Valluri et.,al. \cite{r1}}  \end{figure}
To validate the effectiveness of the JCTFT, a comparative study was conducted using the Inception V3 network \cite{r24,r25} for signal classification. Signals were analyzed using both the JCTFT and the Q-Transform \cite{r22}, and it was found that for signals with low SNRs, the JCTFT significantly outperformed the Q-Transform, particularly in the detection of weak signals where the chirp behavior is critical. Figure 2 illustrates the classification results between the two methods. 

In our present work, we now aim to leverage the same JCTFT technique for our glitch classification pipeline - building upon the proven effectiveness of the JCTFT in gravitational wave signal detection. By utilizing the richer information provided by the 3D time-frequency-chirp spectrograms, we expect to improve the performance of glitch classification, especially for signals with complex frequency evolution.

\section{\label{sec:lev4}Data}
Below we will discuss the strain data used from LIGO's Hanford and Livingston Detectors. We have trained a huge catalogue of data- both glitches and merger signals. The nature of the data and preprocessing are discussed below in the upcoming sub-sections. 
\\The goal would be to process data to the finest detail and reduce it to minimum memory to save the computational cost and time. After the chirp transform we want the data in a format which is very light for the model to train on since it will be working with 1000s of signals and has to be integrated with a live working pipeline later on.
\subsection{Glitches}
The glitches used in this study were provided by R. Girgaonkar and S. D. Mohanty et.,al. \cite{r2}, who identified them in their research. These glitches consist of 708 individual instances of time series strain data, which include attributes such as Time of Arrival ( TOA ), Signal-to-Noise Ratio ( SNR ) and other relevant parameters like Xspacing, starting and ending time, etc. The data were sourced from LIGO's and VIRGO's open-access databases, specifically from the Gravitational Wave Open Science Center ( GWOSC )\cite{r26}. After downloading the data, it was further processed to estimate the Time of Arrival ( TOA ) and the chirp rates $\tau_0$ and $\tau_1$ for each signal ( not to be confused with our linear chirp rate $\gamma$ )- these chirp rates are as discussed in Section 2.B where the context is of 2PN ( second
post-Newtonian ) waveform. Specifically, $\tau_0$ and $\tau_1$ defines the rate at which the waveform’s frequency changes with
time for the zeroth and first orders of frequency, respectively.
This was done by taking 140 hours of LIGO data that had no confirmed merger events or CBC injections.
PSO based matched filtering were done on segments and whichever crossed the thresholds had to be glitches confused as merger events by the matched filtering.
In their work, these chirp rates were used to generate templates that delineated physical and unphysical regions in the signal analysis.

For consistency with our merger signals, which are also 6 seconds in duration, we have clipped each glitch signal to a 6-second window centered around its TOA (shifted around randomly within $\pm1$ later on to ensure temporal robustness). This duration was chosen because it aligns with the typical maximum length of merger signals which generally last for only a few seconds. And the maximum length of glitches could be of a few seconds.
Additionally, this choice helps manage computational costs, as longer signals would result in significantly larger chirp transform outputs - making the analysis more resource-intensive. While we have set this 6-second window as our current standard, future improvements in machine learning models may allow us to reduce this duration even further and have self adjusted time segment lengths potentially enabling the classification of shorter signal segments.

The dataset encompasses a range of SNR values, from a minimum of 9 to a maximum of 14401, and includes glitches recorded during both LIGO Livingston and LIGO Hanford runs. This variety ensures a comprehensive representation of glitch characteristics across different SNR levels and observational settings. 
\subsection{Merger Signals}
The merger dataset used in this work is derived from real gravitational wave strain data obtained from the Hanford (H1) and Livingston (L1) detectors. A total of 215 merger events were selected from the dataset updated in GWOSC as confirmed mergers (until before April 15, 2026).
Since strain from both detectors are treated independently, this results in 430 merger signal instances (215 events × 2 detectors). The SNRs range from anywhere between 4 to around 10.

To increase the size of the training dataset and ensure some robustness to temporal shifts, data augmentation was performed by applying small temporal shifts to each signal. The merger waveform within the six-second analysis window was shifted by a random amount not exceeding $\pm1$ s while preserving the signal morphology and signal-to-noise characteristics. This augmentation approximately doubled the number of available merger samples. 

During preprocessing and volume generation, a small number of signals produced empty or invalid outputs due to unsuccessful API call from GWOSC. These events were excluded from further analysis. In total, 30 augmented samples were discarded, resulting in a final merger dataset consisting of 830 valid merger strain signals and volumes.

The use of augmented real detector data rather than simulated waveforms, allows the model to learn directly from realistic detector noise conditions and astrophysical signal characteristics. Furthermore, the final merger dataset size is comparable to the number of glitch samples used in this study, helping to maintain class balance during training and reducing the possibility of bias towards either the merger or glitch class.
\subsection{Chirp transform and preprocessing}
\begin{figure*}[htbp]
    \centering
    \begin{minipage}{0.4\textwidth}
        \centering
        \includegraphics[width=\textwidth]{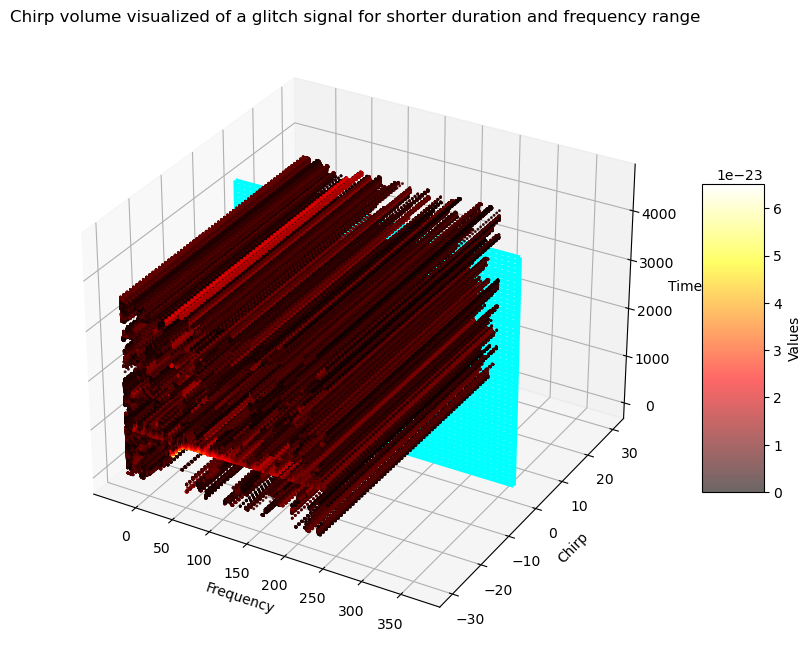}
        \\ (a) 3D chirp volume spectrogram for a glitch signal - sample slice at chirp-rate = 0. 61 such slices to get the 2D time frequency spectrogram. Total 61 slices
    \end{minipage}\hfill
    \begin{minipage}{0.3\textwidth}
        \centering
        \includegraphics[width=\textwidth]{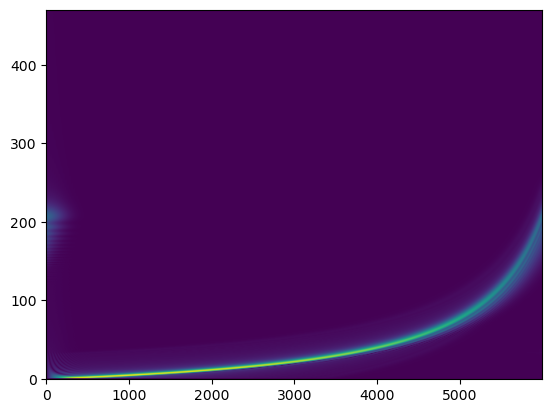}
        \\ (b) 2D time-frequency spectrogram at chirp-rate = 0
    \end{minipage}\hfill
    \begin{minipage}{0.3\textwidth}
        \centering
        \includegraphics[width=\textwidth]{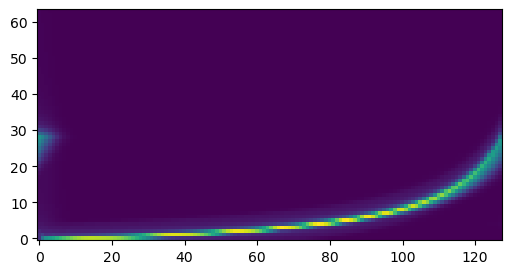}
        \\ (c) Resized time frequency spectrogram slice- resized (471, 6000) to (64,128)
    \end{minipage}
    \caption{The chirp transform and preprocessing: (a), (b), and (c) correspond to the positive and negative domain distributions sliced and resized. First each slice corresponds to (time,frequency) at each chirp hence (471,6000)- this is resized to (64,128). We lose resolution but this is essential for data storage and model processing. Heavy 1.2 GB 3D spectrograms now become just 1.19 MB. (61,471,600) reshaped to (61,64,128)}
    \label{fig:images}
\end{figure*} % expand the images for visibility
To effectively manage and analyze the extensive arrays produced by our chirp transform, we have implemented a strategy that balances data efficiency with analytical precision. 

For each 6-second signal processed through the chirp transform, we generate a large array with dimensions of ( 61, 471, 6000 ). Here, the dimension 61 corresponds to the chirp axis which ranges from -30 to 30 with a step size of 1. The dimension 471 represents the frequency range, spanning from 30 Hz to 500 Hz. The reason we pick this frequency range is because 30 Hz is the lower cutoff of what the PSO based matched filtering had searched over ( and most physically plausible merger signals fall above this threshold ) though the upper cutoff for the detector is 700Hz,  most of the physically possible inspiral merger signals fall within the 500 Hz range ( with exception of some very high mass binaries ). To keep computational cost at a reasonable level we stop with 500 Hz cutoff since our model which we later design is adept even for this frequency range input. Finally, the dimension 6000 corresponds to the time samples, which are derived based on a Nyquist sampling rate of 1000 \cite{r34} to accommodate the 500 Hz cutoff ( minimum downsampling rate= 2 * frequency cutoff. In our case 2 * 500 = 1000 ). This results in a complex datatype array that occupies approximately 1.2 GB of storage. 

Handling such voluminous datasets for thousands of signals presents considerable challenges in terms of both storage and processing capacity. To address these challenges, we have devised a method to reduce the size and complexity of the data while retaining its analytical value. 

To make the dataset tractable for large-scale analysis, we reduce the dimensionality of each chirp-volume while preserving its dominant spectro-temporal structure. Each of the 61 chirp-rate slices is treated as an individual two-dimensional time-frequency spectrogram and resized from \(471 \times 6000\) to \(64 \times 128\) using linear interpolation. This operation is analogous to image downsampling and preserves the overall morphology of the spectrogram while significantly reducing the number of stored data points.

Following resizing, each spectrogram slice is independently normalized to the range \([0,1]\). This ensures a consistent dynamic range across all chirp slices and prevents variations in absolute signal amplitude from dominating subsequent analysis.

The processed chirp-volume is therefore represented as a tensor of dimensions

\[
(61,64,128),
\]

corresponding to the chirp-rate, frequency, and time axes respectively. The resulting tensors are stored as NumPy arrays, reducing the storage requirement of a single signal from approximately 1.2 GB to approximately 1.9 MB. This reduction enables efficient storage and processing of large signal catalogues while retaining the key characteristics of the original chirp-transform representation.

Figure 3 illustrates the complete preprocessing pipeline adopted in this work. The resulting chirp-volume representation provides a compact and information-rich description of each signal, enabling efficient storage, processing and large-scale machine learning analysis while retaining the characteristic structures required for distinguishing merger signals from glitches.

\section{Principal Component Analysis ( PCA )}
In this study, we utilized Principal Component Analysis ( PCA ) \cite{r18} to reduce the dimensionality of the features extracted from the normalized histogram distributions representing the signals. This was a primary test to visualize the differences between glitch and merger signals in a lower-dimensional space, which would make patterns more discernible and facilitate a better understanding of the underlying structure in our data.\\   PCA is a statistical technique that transforms high -dimensional data into a set of orthogonal axes known as principal components, capturing the directions of maximum variance. By projecting our high-dimensional feature space onto two principal components, we could create a 2D representation that highlights the most significant patterns and differences within the dataset. Specifically, we computed PCA1 and PCA2. PCA1 corresponds to the direction of maximum variance and represents the first principal component derived from a linear combination of various histogram bins from both the positive and negative chirp domains. Similarly, PCA2 captures the second most significant variance, providing additional context and further distinguishing features that are orthogonal to PCA1.

The visualization of the signals along these principal components, as illustrated in Figure~\ref{fig
}, revealed a notable separation between glitch and merger signals. Most signals were clearly clustered into distinct groups, corresponding to either glitches or mergers with some instances of overlap. This clear separation indicates that the statistical characteristics of glitches and mergers are sufficiently distinct to be differentiated based on their principal components. The PCA analysis effectively highlighted these differences, supporting the robustness of our preprocessing steps.

However despite the clear separation, some overlap between the two signal types remained. This overlap can be attributed to the inherent variability within the signals and potential limitations in the feature extraction process. To address these challenges and achieve more accurate classification, we employed Convolutional Neural Networks ( CNNs ) along with GRU and attention mechanism. CNNs are particularly adept at learning complex patterns and relationships in the data, achieving higher accuracy than PCA alone. While PCA serves as a valuable tool for initial exploration and visualization, advanced models like CNNs are essential for achieving superior classification performance. \cite{r19}

The precise linear combinations that define PCA1 and PCA2 are detailed in the supplementary materials available on our GitHub repository. The script contains an exhaustive list of contributing factors and weights used to compute the principal components, which is not possible to give as an equation in the context of this paper. But as explained earlier, they are linear combinations of the bins data from both positive and negative chirp distributions such that the power distribution ( or counts of bins here ) has the most variance possible.

In summary, PCA proved to be an effective method for dimensionality reduction and visualization, revealing distinct patterns that differentiate glitch and merger signals. While PCA provided valuable insights into the data, the application of CNNs with GRU Attention Mechanism in the next section enhanced our classification accuracy, demonstrating the complementary role of these techniques in our analysis.

\begin{figure}[htbp] \centering \fbox{\includegraphics[width=\linewidth]{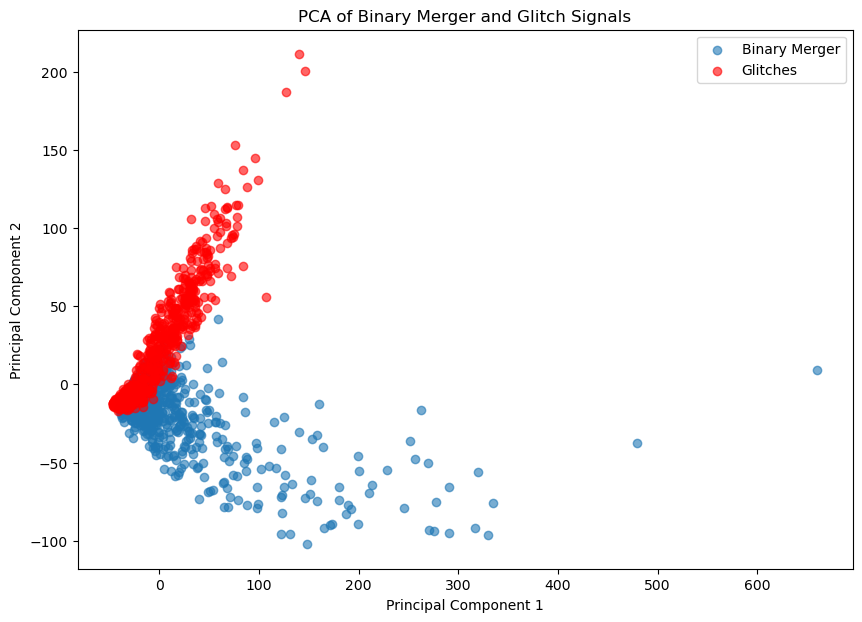}} \caption{Principal Component Analysis ( PCA ) of the normalized distributions. PCA1 represents the linear combination of histogram bins that captures the most variance while PCA2 captures the second most variance in the power distribution.  The plot shows a clear separation between glitches and mergers, with some overlap indicating the need for advanced classification techniques. These principal components help visualize the differences between signal types and support the robustness of our classification approach.} \label{fig
} \end{figure}
\section{Network model}
We employ a hybrid Convolutional Neural Network- Gated Recurrent Unit (CNN- GRU) architecture \cite{r13}\cite{r67} with an attention mechanism \cite{r68} for this classification task. Deep neural networks are particularly suitable for analyzing chirp-volume spectrograms because they can automatically learn discriminative features from high-dimensional data. The input to the model consists of chirp-volume tensors of dimensions $(61,64,128)$, where the first dimension corresponds to the chirp-rate axis and the remaining dimensions correspond to the frequency and time axes respectively resized. Each chirp-rate slice is treated as a two-dimensional spectrogram and processed independently by a convolutional feature extractor.
\begin{figure*}[htbp]
    \centering
    \includegraphics[width=\textwidth]{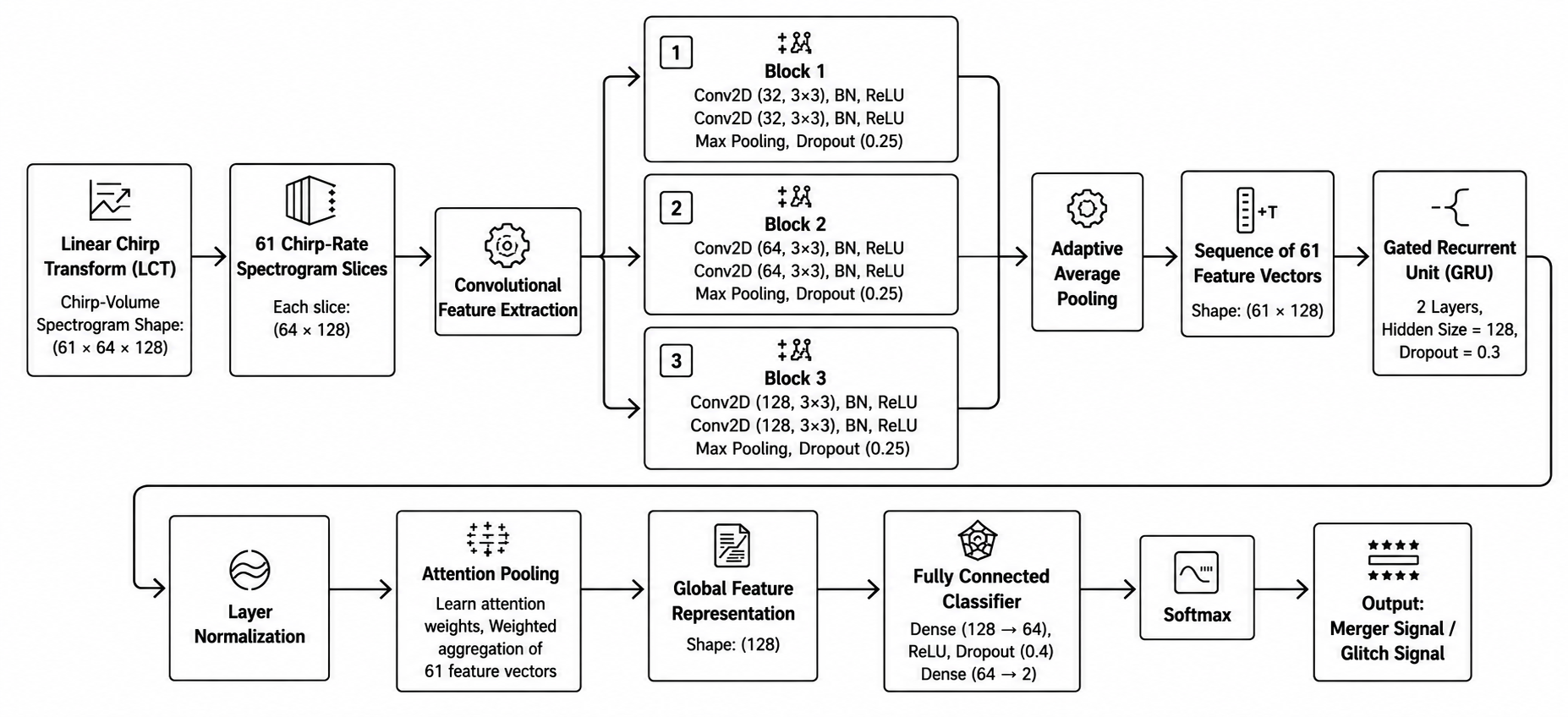}\hfill
    \caption{CNN-GRU based Attention architecture for Glitch Veto. }
\end{figure*}
The convolutional component of the network consists of three successive blocks. The first block employs two convolutional layers with 32 filters, batch normalization, Rectified Linear Unit (ReLU) activations \cite{r15}, max-pooling operations \cite{r14} and dropout regularization \cite{r69}. The second and third blocks increase the number of filters to 64 and 128 respectively. Throughout the network, small \(3\times3\) convolution kernels \cite{r16} are used to capture fine-scale structures within the spectrograms. These layers progressively extract increasingly complex representations while reducing the spatial dimensionality of the input.

Following the convolutional stages, adaptive average pooling \cite{r70} is applied to each chirp-rate slice, producing a compact 128-dimensional feature vector. The resulting sequence of 61 feature vectors is then processed using a two-layer Gated Recurrent Unit (GRU) network. The GRU models correlate between neighbouring chirp-rate slices and learns how signal morphology evolves across the chirp domain. Layer normalization \cite{r71} is subsequently applied to improve training stability and convergence.

Rather than relying solely on the final recurrent state, we incorporate an attention pooling mechanism. The attention layer assigns a learnable weight to each chirp-rate slice and computes a weighted combination of all GRU outputs. This allows the network to focus on the most informative chirp regions while suppressing less relevant features. The resulting representation is passed through a fully connected classification head consisting of a dense layer with 64 neurons, ReLU activation and dropout regularization.

The final output layer contains two neurons corresponding to the merger and glitch classes. A softmax activation function \cite{r17} converts the output into a probability distribution over the two classes. To improve robustness against class imbalance and difficult training examples, we employ focal loss during training. Optimization is performed using the AdamW optimizer \cite{r72} together with cosine annealing learning-rate scheduling \cite{r73}, while mixed-precision training \cite{r74}, gradient clipping \cite{r75}, and early stopping \cite{r76} are utilized to improve computational efficiency and convergence stability.

The combination of convolutional feature extraction, recurrent sequence modelling and attention-based aggregation enables the proposed architecture to effectively distinguish between merger signals and glitches. By simultaneously exploiting local spectrogram structures and long-range correlations across the chirp domain, the model provides a robust framework for gravitational-wave signal classification.

The quantitative results are discussed in the following subsection

\section{Results}
The proposed CNN- GRU- Attention model was evaluated on an independent validation set comprising 308 signals, corresponding to approximately 20\% of the available dataset. Training was done on the remaining 80\% of the available dataset - 1230 signals. It was ensured that the glitches and mergers signals were randomly and equally distributed in train and validation sets to avoid class imbalance.
The network converged after 57 epochs through early stopping and achieved a best validation accuracy of 95.45\%.

To optimize classification performance, a threshold tuning procedure was performed on the validation set. The optimal threshold was found to be 0.44, yielding an overall classification accuracy of approximately 96\%. At this operating point, both merger and glitch classes achieved comparable precision and recall values of approximately 95-96\%, indicating balanced performance across the two classes.

The confusion matrix shown in Figure 6 demonstrates the strong agreement between the predicted and true labels. Out of 308 validation samples, only 13 were misclassified. The resulting macro-averaged precision, recall, and F1-score were all approximately 0.96, demonstrating that the model generalizes well and is not biased toward either class.

\begin{figure}[htbp]
\centering
\fbox{\includegraphics[width=\linewidth]{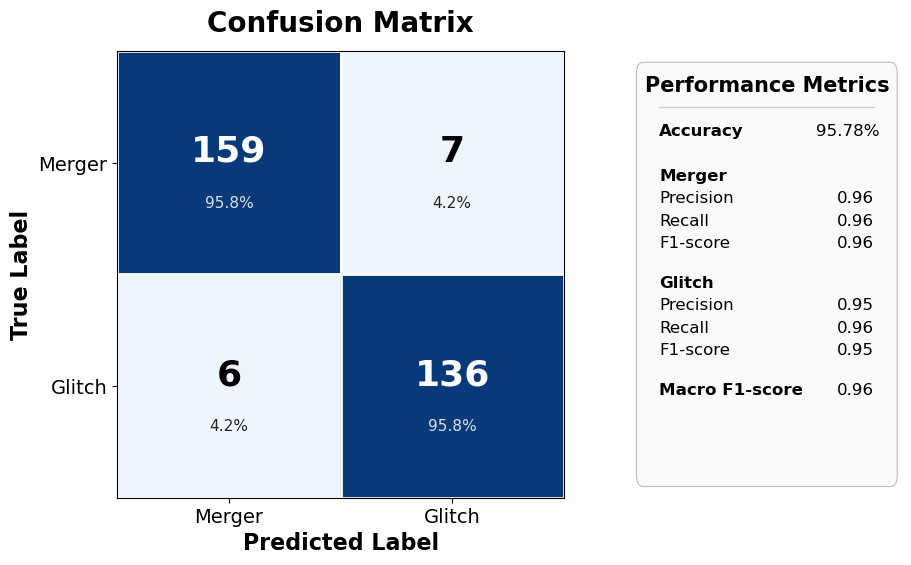}}
\caption{Validation performance on test dataset - test size $\approx$ 0.2 of the total dataset or 308 signal samples. We could clearly see only 13 signals (6 glitches and 7 mergers) misclassified out of 308. All in all the precision and recall consistently cross 0.95 and is a highly accurate model for current distinction between glitches and merger signals}
\end{figure}
These results indicate that the chirp-volume representation generated by the Linear Chirp Transform contains highly discriminative information for distinguishing merger signals from glitches. The combination of convolutional feature extraction, recurrent sequence modelling and attention-based aggregation enables the network to effectively exploit both local spectrogram features and global chirp-domain correlations, leading to robust classification performance.

\section{Conclusion and future possibilities}
We have presented a machine learning-based method of the chirp transform to differentiate glitches from merger signals. Our approach is intended to be a part of pipelines in the context of merger signals after a signal has passed the matched filtering stage and is labelled as a candidate event - to validate if it is a glitch or not. We trained the model on NVIDIA GeForce GTX 1660 Ti GPU card from personal laptop with memory allocated of 4GB. You could train on CPUs as well- but in batches and the training would take a few hours or more.
It took around 32.7 seconds to perform the chirp transform, resize and store the desired numpyvolumes. The stored trained model takes 30 ms to classify if a given signal is a glitch or not, when we input the resized chirp volume. 
\\This gives a highly accurate model for differentiating glitches and signals in the context of inspiral merger signals. Though trained for LIGO detectors, the logic of signals separating out in the chirp domain could be used for new age and advanced future detectors given the merger signal has a varying frequency profile and chirp transform to capture the features better than traditional fourier transforms. As new types of glitches are discovered they could be included to train in the model. Further analysis could be done to infer more about the applications of the chirp transform for signal processing. More could be inferred from deriving the transform for higher chirp rates or fractional chirp rates. \\  Getting an exact solution is a daunting task but estimates for specific signal processing purposes could be made. One could get well fitted analytical solutions; thereby the inference one could get from various signals could be better leveraged. The Chirp transform has already shown its usefulness in other domains like Radio Detection and Ranging ( RADAR ) and other signal processing tasks \cite{r43,r44} which have varying frequency profiles.  Improvement on solutions of other polynomial chirp rates would contribute to domains that have specific waveform models.
\\Furthermore it is possible to have better data storage methods for the heavy 3D spectrograms so that one could gain more information about a given signal. When this pipeline is included with the best computing resources available - one would get a quicker performance since the model is fast on conventional computers with minimum resources as well.
\\To summarize, we have a highly accurate pipeline to differentiate glitches from genuine inspiral merger signals. We can use the ideology to study other types of glitches like gamma-ray pulsars, fast radio bursts, etc and test our methodology.
%All the scripts and data could be found in our github repository- \hyperref[https://github.com/Arut123/Glitch-veto-GW-using-chirp]{}{{https://github.com/Arut123/Glitch-veto-GW-using-chirp}

\section{Acknowledgement}
This work was undertaken under MITACs Globalink GRI'24 fellowship conferred upon Arutkeerthi supervised by Prof. Sree Ram Valluri. We acknowledge valuable suggestions given by Dr. Soumya Mohanty and Raghav Girgaonkar of University of Texas Rio Grande Valley who gave a vision and kickstart to this work by their recent research, sharing the data and suggesting an application of the chirp transform. We also acknowledge Dr.Martin Houde of Western University,Ontario in playing a key role while developing the Chirp Transform. We thank the authors of Tensorflow, Numpy, Matplotlib, PyCBC, GWPy, Scipy and Pandas which we used to complete our work,model and visualise in a rapid manner \cite{r47,r48,r49,r50,r51,r52,r53,r54}.\\\\\textbf{Data Availability} - All the scripts and data used could be found in our github repository- \href{https://github.com/Arut123/Glitch-veto-GW-using-chirp}{https://github.com/Arut123/Glitch-veto-GW-using-chirp}
\bibliography{references}

\end{document}